\documentclass[manuscript]{aastex631}
\usepackage{array}
\usepackage{float}
\usepackage{xcolor}
\usepackage{enumitem}
\usepackage{booktabs}
\usepackage{amsmath}
\usepackage{amssymb}
\usepackage{libertine}
\usepackage{multirow}
\usepackage{epstopdf}

\newcommand{\B}{$\beta$}
\newcommand{\A}{$\alpha$}
\newcommand{\BG}{$\beta \gamma$}
\newcommand{\G}{$\gamma$}
\newcommand{\BD}{$\beta \delta$}
\newcommand{\BGD}{$\beta \gamma \delta$}
\newcommand{\GD}{$\gamma \delta$}
\newcommand{\D}{$\delta$}

\shorttitle{Correlation between Solar Energetic Particles and Active Regions}
\shortauthors{Marroquin et al.}
\graphicspath{{./}{figures/}}

\begin{document}

\title{Statistical Study of the Correlation between Solar Energetic Particles and Properties of Active Regions}

\correspondingauthor{Viacheslav Sadykov}
\email{vsadykov@gsu.edu}

\author[0000-0002-3364-7463]{Russell D. Marroquin}
\affiliation{Department of Physics, University of California San Diego, La Jolla, CA 92093, USA}
\affiliation{Physics \& Astronomy Department, Georgia State University, Atlanta, GA 30303, USA}

\author[0000-0002-4001-1295]{Viacheslav Sadykov}
\affiliation{Physics \& Astronomy Department, Georgia State University, Atlanta, GA 30303, USA}  

\author[0000-0003-0364-4883]{Alexander Kosovichev} 
\affiliation{Physics Department, New Jersey Institute of Technology, Newark, NJ 07102, USA}
\affiliation{NASA Ames Research Center, Moffett Field, CA 94035, USA}

\author[0000-0003-4144-2270]{Irina N. Kitiashvili}
\affiliation{NASA Ames Research Center, Moffett Field, CA 94035, USA}

\author{Vincent Oria}
\affiliation{Computer Science Department, New Jersey Institute of Technology, Newark, NJ 07102, USA}

\author[0000-0003-2846-2453]{Gelu M. Nita}
\affiliation{Physics Department, New Jersey Institute of Technology, Newark, NJ 07102, USA}

\author[0000-0002-2858-9625]{Egor Illarionov}
\affiliation{Department of Mechanics and Mathematics, Moscow State University, Moscow, 119991, Russia}
\affiliation{Moscow Center of Fundamental and Applied Mathematics, Moscow, 119234, Russia}

\author{Patrick M. O’Keefe}
\affiliation{Computer Science Department, New Jersey Institute of Technology, Newark, NJ 07102, USA}

\author{Fraila Francis}
\affiliation{Computer Science Department, New Jersey Institute of Technology, Newark, NJ 07102, USA}

\author{Chun-Jie Chong}
\affiliation{Computer Science Department, New Jersey Institute of Technology, Newark, NJ 07102, USA}

\author{Paul Kosovich}
\affiliation{Physics Department, New Jersey Institute of Technology, Newark, NJ 07102, USA}

\author[0000-0003-3196-3822]{Aatiya Ali}
\affiliation{Physics \& Astronomy Department, Georgia State University, Atlanta, GA 30303, USA}

\begin{abstract}

The flux of energetic particles originating from the Sun fluctuates during the solar cycles. It depends on the number and properties of Active Regions (ARs) present in a single day and associated solar activities, such as solar flares and coronal mass ejections (CMEs). Observational records of the Space Weather Prediction Center (SWPC NOAA) enable the creation of time-indexed databases containing information about ARs and particle flux enhancements, most widely known as Solar Energetic Particle events (SEPs). In this work, we utilize the data available for Solar Cycles 21-24, and the initial phase of Cycle 25 to perform a statistical analysis of the correlation between SEPs and properties of ARs inferred from the McIntosh and Hale classifications. We find that the complexity of the magnetic field, longitudinal location, area, and penumbra type of the largest sunspot of ARs are most correlated with the production of SEPs. It is found that most SEPs ($\approx$60\%, or 108 out of 181 considered events) were generated from an AR classified with the 'k' McIntosh subclass as the second component, and these ARs are more likely to produce SEPs if they fall in a Hale class containing $\delta$ component. The resulting database containing information about SEP events and ARs is publicly available and can be used for the development of Machine Learning (ML) models to predict the occurrence of SEPs.

\end{abstract}


\section{Introduction}\label{sec:intro}

    Solar Energetic Particle (SEP) events are widely known as particle flux enhancements measured by near-Earth satellites. They are among the most dangerous transient phenomena of solar activity because of their negative impacts including health risks for astronauts and airline crews and passengers, damages to satellites and aircraft, radio wave disturbances, power grid disruptions, etc. \citep{Kataoka2018,Martens2018}. Prediction of SEP events before their occurrence could help diminish their impacts by taking measures ahead of time.

    The use of Machine Learning (ML) to predict SEP events has grown significantly in the past years \citep{Sadykov2021,Torres2022,Kasapis2022}, in part due to the increase in the readiness of data related to these events. As an example, in the recent review of 36 SEP prediction models by \citet{Whitman2022}, 10 models involved ML. Because the performance of ML may be proportional to the quality and availability of data \citep{Goodfellow2016}, we place significant importance on the development of databases with information about SEP events and Active Regions (ARs) expanded in time. ARs are sunspots or sunspot groups that represent regions of strong magnetic fields on the solar surface. They are the primary sources of solar flares and coronal mass ejections (CMEs), and consequently, the SEPs \citep{Reames2021,Toriumi2017}. Therefore, it is important to study the link between SEP productivity and the basic properties of ARs, which include their location on the solar disk, their areas, various magnetic field characteristics, and other measures of AR structure and complexity. Probably the longest available observational measures of the AR structure and complexity are the McIntosh \citep{McIntosh1990} and Hale classifications \citep{Hale1919}. In these classifications, higher classes are assigned to larger and more complex ARs, which are typically more productive in terms of flares, CMEs, and SEPs.

    The correlations between SEPs and other solar transient phenomena (flares and CMEs) and McIntosh and Hale classifications were previously studied in several works. For example, \cite{Bronarska2017} studied 84 SEP events recorded during the Solar and Heliospheric Observatory (SOHO) spacecraft era (1996-2014) with ARs characterized by the McIntosh classification and found that the most energetic SEPs are ejected only from the associated ARs having a large and asymmetric penumbra. \cite{Toriumi2017} analyzed the Hale classification of ARs that govern large solar flares and eruptions observed between May 2010 and April 2016. \cite{McCloskey2016} studied the flaring rates and the evolution of ARs in terms of the McIntosh classes using data for Solar Cycle 22 (SC22). Their results supported the hypothesis that injection of magnetic energy by flux emergence, which results in an increase in the AR class in the McIntosh system, leads to a higher frequency and magnitude of flare events and, thus, higher SEP event production. The operational flare and SEP daily probabilistic forecasting efforts at the Space Weather Prediction Center (SWPC) at the National Oceanic and Atmospheric Administration (NOAA) also rely on the AR classes \citep{Bain2021}

    The studies mentioned above confirmed the importance of the AR classification and the need for long cross-cycle studies of AR properties and related transient activity. The primary goals of this paper are (1) to develop a homogeneous 40-year-long period data set of Hale and McIntosh classes of active regions and associated SEP events spanning from December 1981 to December 2021, which covers the declining phase of SC21, SC22-24, and the rising phase of SC25, and (2) to perform a direct statistical study of the correlations between SEPs and the Hale and McIntosh classes of ARs using these data. The catalogs utilized in this study are the Solar Region Summary (SRS) records available from the SWPC NOAA\footnote{\url{ftp://ftp.swpc.noaa.gov/pub/warehouse/}} starting from 1996, the United States Air Force (USAF) records of AR classes\footnote{\url{https://www.ngdc.noaa.gov/stp/space-weather/solar-data/solar-features/sunspot-regions/usaf_mwl/}} available from 1981 until 2017, and the catalog of the solar energetic particle events affecting the Earth maintained by NOAA\footnote{\url{https://umbra.nascom.nasa.gov/SEP/}} with the information about proton events available since 1976. The paper is structured as follows. Section~\ref{sec:classes} describes the McIntosh and Hale classifications of ARs. Section~\ref{sec:data} describes the catalogs used in this study and related data preparation steps. Section~\ref{sec:analysis} highlights the results of the statistical analysis of the association of AR classes and properties with SEPs, followed by the summary of our findings in Section~\ref{sec:summary}. The generated homogeneous data set of AR properties spanning from 1981 until 2021 is publicly available at the Solar Energetic Particle Prediction Portal (SEP$^3$) webpage\footnote{\url{https://sun.njit.edu/SEP3/datasets.html}}.

\section{Description of Active Region Classification} \label{sec:classes}

    \subsection{Hale Classification of Active Regions} \label{sec:Hale}

    The Mount Wilson (or Hale) classification system for sunspot groups put forward by \citet{Hale1919} has been used for nearly a century. This type of magnetic classification provides a simple way to describe the configuration of the magnetic flux and sunspots in an AR \citep{Jaeggli2016}. According to this classification, $\alpha$ is a unipolar sunspot group configuration, $\beta$ is a distinct bipolar configuration with opposite magnetic polarities, $\gamma$ is a complex configuration with irregular distribution of polarities, and $\delta$ is a configuration with a sunspot umbra that contains opposite magnetic polarities separated by less than $2^\circ$ within one penumbra. When appropriate, the classification can include combinations of the primary classes. For example, $\beta \gamma$ is a complex bipolar configuration with more than one continuous line connecting the opposite polarities. Besides that, if the sunspot group configurations contain one or more $\delta$ spots, the $\delta$ class is added to the $\beta$-$\gamma$ classification \citep{He2021}. Here we summarize the classes and subclasses presented in the USAF data set:

    \newlist{contract}{enumerate}{10}
    \setlist[contract]{label*=\arabic*.}
    \setlistdepth{10}
        
    \begin{contract}
        
    \item ALPHA (\A). A single sunspot, or a unipolar sunspot group, around which the distribution of plage is fairly symmetrical. Magnetic field measurements show that the unipolar groups are often accompanied by an area of opposite polarity in which sunspots are not visible.
        \begin{contract}
        \item ALPHA p (\A). The magnetic field polarity in and around the spot(s) corresponds to the polarity of the leader spots in the same hemisphere for the current solar cycle. The spot(s) and adjacent plage are followed by an elongated area of plage or faculae of the opposite polarity (used in the USAF classification only).
            
        \item ALPHA f (\A). The magnetic field polarity in and around the spot(s) corresponds to the polarity for the trailer spots in the same hemisphere for that cycle. The spot(s) and adjacent plage are preceded by an elongated area of plage or faculae of the opposite polarity (USAF only).
        \end{contract}
        
    \item BETA (\B). A bipolar group in which magnetic field strengths and spot areas indicate a balance between the leader and trailer spots. The polarities show a clear separation.
        
        \begin{contract}
            \item BETA p (\B). A bipolar group, in which the magnetic field strengths and spot areas indicate that the leader spot is dominant (USAF only).
            
        \item BETA f (\B). A bipolar group, in which the magnetic field strengths and spot areas indicate that the trailer spot is dominant (USAF only).
        \end{contract}
    
    \item BETA-GAMMA (\BG). A spot group that has \B \space characteristics, but is lacking a well-defined dividing line between regions of opposite polarity. This class includes cases in which spots of the opposite or ``wrong'' polarity accompany the leader or trailer regions.
        
    \item GAMMA (\G). A spot group in which the polarities are completely intermixed.
        
    \item BETA-DELTA (\BD). A spot group, which has \B \space characteristics, but has umbrae of opposite polarity inside the same penumbra.
        
    \item BETA-GAMMA-DELTA (\BGD). A spot group, which has \BG \space characteristics, but has umbrae of opposite polarity inside the same penumbra.
        
    \item GAMMA-DELTA (\GD). A spot group, which has \G \space characteristics, but has umbrae of opposite polarity inside the same penumbra.
        
    \item DELTA (\D). A sunspot group with umbra having opposite polarities within a penumbra \textbf{that} spans less than two heliographic degrees.
        
    \end{contract}

    \subsection{McIntosh Classification of Active Regions} \label{sec:McI}

    The McIntosh classification scheme was originally developed by \citet{Cortie1901} and later modified and expanded to include a wider range of parameters by \citet{McIntosh1990}. It describes the white-light structure of sunspot groups and is composed of 60 allowed classification combinations derived from 17 different parameters \citep{McCloskey2016}. The general form of the McIntosh classification is \emph{Zpc}, where \emph{Z} is the modified Zurich class, \emph{p} is the type of largest spot, and \emph{c} is the degree of compactness in the interior of the group. A more detailed description of the classification scheme based on the USAF database documentation is provided below.

        \subsubsection{Modified Zurich Class -- Z}
        The modified Zurich classes are defined on the basis of whether penumbra is present, how the penumbra is distributed, and by the extent of the group \citep{McIntosh1990}. In contrast to the original Zurich definitions, a judgment of complexity is not required. There are seven classes in this component of the system described below.
        
         \begin{contract}
            \item[$A$] \quad Unipolar group with no penumbra with the total extent (normally) of less than 3 heliographic degrees.
            
            \item[$B$] \quad Bipolar group of spots with no penumbra; the length is (normally) 3 heliographic degrees or greater.
            
            \item[$C$] \quad Bipolar group of spots when only spots of one polarity have penumbra, usually the spots at one end of an elongated group.
            
            \item[$D$] \quad Bipolar group when spots of both polarities have penumbra. The group length does not exceed 10 heliographic degrees.
            
            \item[$E$] \quad Bipolar group when spots of both polarities have penumbra.  The group length is greater than 10 but less than or equal to 15 heliographic degrees.
            
            \item[$F$] \quad Bipolar group when spots of both polarities have penumbra. The group length exceeds 15 heliographic degrees.
            
            \item[$H$] \quad Unipolar group of spots with penumbra. The principal spot is usually the leader spot remaining from an old bipolar group.
    
        \end{contract}

        \subsubsection{Penumbra of Largest Spot -- p}
        
        The type of largest spot in a sunspot group can be described by a combination of type, size, and symmetry of penumbra and umbrae within a given penumbra \citep{McIntosh1990}. There are six classes in this component described below.
        
        \begin{contract}
            \item[$x$] \quad  No penumbra.
            
            \item[$r$] \quad Rudimentary or incomplete irregular penumbra. It is brighter than a mature penumbra and has a mottled or granular (not filamentary) fine structure.
            
            \item[$s$] \quad Small symmetric penumbra. Mature, dark, circular, or elliptical penumbra with a filamentary fine structure; the diameter across the penumbra is 2.5 heliographic degrees or less. This class includes penumbrae that appear elliptical due to the effect of geometric foreshortening. Symmetric penumbra usually contains either a single umbra or a compact cluster of umbrae near the sunspot center.
            
            \item[$a$] \quad Small asymmetric penumbra. Mature, dark, irregular (clearly not circular or elliptical) penumbra with filamentary fine structure; the diameter across the penumbra is 2.5 heliographic degrees or less. The asymmetry is ``real'', not just due to foreshortening effects. An asymmetric penumbra usually contains two or more umbrae scattered within it.
            
            \item[$h$] \quad Large symmetric penumbra. It has the same characteristics as a small symmetric(s) penumbra; the diameter across the penumbra is greater than 2.5 heliographic degrees (normally corresponding to an area greater than about 250 millionths of the solar hemisphere).
            
            \item[$k$] \quad Large asymmetric penumbra. It has the same characteristic as a small asymmetric (a) penumbra; the diameter across the penumbra is greater than 2.5 heliographic degrees (normally corresponding to an area greater than about 250 millionths of the solar hemisphere).
            
        \end{contract}

        \subsubsection{Sunspot Distribution -- c}

        This component of the three-letter classification indicates the density of an internal spot population in a sunspot group. A ranking of spot distribution in the interior of a sunspot group gives additional information about the area of the group and the potential presence of strong spots near the line of polarity inversion lying between the principal leader and follower spots \citep{McIntosh1990}.
    
        \begin{contract}
            \item[$x$] \quad Undefined for a single sunspot or unipolar spot group.
            
             \item[$o$] \quad Open. Few, if any, spots between the leader and trailer spots. Any interior spots are very small umbral spots or pores.
             
             \item[$i$] \quad Intermediate. Many spots lie between the leading and trailing portions of the group, but none of them possesses mature penumbra.
             
             \item[$c$] \quad Compact. The area between the leading and trailing ends of the spot group is populated with many strong spots, with at least one interior spot possessing mature penumbra. An extreme case has the entire spot group enveloped in one continuous penumbral area.
            
        \end{contract}

    A summary of the McIntosh classification is given in Table~\ref{table:2} The total number of possible classes is 60. It is important to note that not every combination of the components mentioned above is permitted.

    \begin{table}[!ht]
    \begin{tabular}{| c | c | c | c |} 
     \hline
     Class  & Penumbra: Largest Spot & Distribution & Number of Configurations \\ [1ex]
     \hline\hline
     A & x & x & 1 \\ 
     \hline
     B & x & o,i & 2 \\
     \hline
     C & r,s,a,h,k & o,i & 10 \\
     \hline
     D,E,F & r & o,i & 6 \\
     \hline
     D,E,F & s,a,h,k & o,i,c & 36 \\
     \hline
     H & r,s,a,h,k & x  & 5 \\
     \hline
     Total allowed types & -- & -- & 60\\  
     \hline
    \end{tabular}
    \caption{McIntosh Class Configurations of Sunspot Groups.}
    \label{table:2}
    
    \end{table}

\section{Data preparation} \label{sec:data}

    \subsection{Homogeneous Data Set of Solar Active Regions}\label{sec:ARdata}

        The USAF dataset of active region records contains information about AR classes from December 1981 to December 2017. The currently-maintained SWPC NOAA SRS records range from January 1996 to the current time. To maximize the availability of data for this statistical analysis, we combine the USAF and SWPC NOAA AR catalogs into a continuous AR database covering the 40-year period, from December 1981 to December 2021. However, the catalogs are not entirely consistent in the way the AR classes are reported. For example, SWPC NOAA SRS records contain information about ARs once per day at 00:00\,UT, while the USAF dataset can contain several records of the same AR throughout the day from different observing sites at different times. Moreover, many solar energetic particle events are originating from the regions close to the western limb where the information about AR is either not available or ambiguous. Therefore, we have performed certain steps toward the homogenization of the data sets, namely bringing the records from the USAF data set into a form compatible with the current SWPC NOAA SRS reporting.

        The USAF catalog files were held separately for each year and in a text file format. Relevant information to perform this statistical analysis was acquired line by line and character by character, following the documentation provided for this data catalog. In order to keep consistency with the SWPC NOAA records and the accuracy of its contents, these records were edited in accordance with the documentation provided by the USAF. The years provided in a \emph{YY} format were changed to \emph{YYYY} format so that the dates followed the consistent format \emph{YYYY-MM-DD hour:min:sec}. According to the documentation, the location of an observed AR is given respectively by six characters in the following order: E or W for East or West, two integers for longitude given as central meridian distance, N or S for North or South, and two integers for heliographic latitude. The negative sign was added for the South and East longitudes.

        The records from the USAF came from four different observational stations, such as the Mount Wilson Observatory and the Boulder Observatory among others. Following the documentation, the area of each AR is given in millionths of a solar hemisphere. It was noticed that some entries of the area were equal to zero, while additional entries for the same AR from other observatories had nonzero areas. We decided that if the area of an AR was equal to zero while being recorded as nonzero by any other observatory, such an AR would take its respective and most recent nonzero area value, independently from which observatory the information was obtained. This method was carried out through an algorithm by first sorting the respective dataset by AR number, date, and time of observation, which resulted in groups of ARs that ascended based on the parameters just mentioned. As our algorithm iterated through each entry if an entry with a value of area equal to zero was reached, it inspected the previous or next most recent records of the respective AR. If the algorithm found a nonzero value for the same AR, it replaced the zero value and moved on to the next entry.

        According to the documentation, the USAF dataset followed the Hale classification scheme described in Section \ref{sec:Hale}. A variety of inconsistencies were found throughout the catalogs for each year, and such inconsistencies were updated accordingly. Hale classifications were given as single letters with up to four character spaces, depending on the sunspot group configuration, and were changed accordingly. For instance, \emph{A} was replaced by \A, \emph{B} was replaced by \B, \emph{BGD} was replaced by \BGD, and so on for every entry. Subclasses of \B \space and \A \space (items 1.1, 1.2, 2.1, and 2.2 in Section \ref{sec:Hale}) were designated as \B \space and \A, respectively. Following the sorting method based on the AR number, date, and time, an algorithm iterated through every entry. When a Hale class was considered ambiguous, meaning that its magnetic type was unclear, we inspected the previous or next most recent records of the respective AR. If the algorithm found an appropriate magnetic type for the same AR, it replaced the Hale class in question and moved on.

        The McIntosh classification scheme described in Section \ref{sec:McI} is inferred from \citet{McIntosh1990} and the USAF documentation. The three-component McIntosh classes given by three characters were extracted one by one from the USAF datasets. Inconsistencies in these records were divided into two categories: ``ambiguous'' and ``unambiguous''. The McIntosh classes were designated ambiguous when one of the three components was missing, and this missing component had more than one allowed class outlined in Table \ref{table:2}. For instance, the McIntosh classification \emph{'F I'} was labeled as ambiguous because the penumbra type of the largest spot (second component) was not specified, and based on table \ref{table:2}, several different classes in the second component are allowed to be paired with \emph{'F'} class (first component) and \emph{'I'} distribution of sunspots (third component). In addition, several McIntosh configurations contained one or more components paired with one or more disallowed components. To illustrate, the McIntosh configurations starting with the \emph{'H'} class as the first component and any penumbra type of the largest spot  (listed in Table \ref{table:2}) were considered as "ambiguous" because five different classes are allowed to take its place as the second component. Nevertheless, any ARs classified with \emph{'H'} for the first component whose third component is inconsistent can be considered "unambiguous" because only the \emph{'X'} class is allowed in place of such inconsistency. As a result, the McIntosh Classification \emph{'HXO'} was labeled as "ambiguous" because of the following reasons: 1) The \emph{'H'} class is not allowed to be paired with the second component \emph{'X'} or the third component \emph{'O'} as defined in Table \ref{table:2}. 2) Although the third component can be unambiguously edited to be \emph{'X'}, there exist five different allowed classes for the second component, leaving no unambiguous choice for such a component.

        In contrast to the ``ambiguous'' McIntosh classifications, an ``unambiguous'' classification refers to a McIntosh configuration with specifically one or more missing components, which can be unambiguously added to the respective record. A caveat of the last sentences is that the word ``specifically'' is important because, for example, the McIntosh classification \emph{'AXO'} could easily be considered as ``unambiguous'' since the third component can be unambiguously replaced with \emph{'X'} to form \emph{'AXX'}. Nevertheless, according to Table \ref{table:2}, the first component \emph{'A'} can be edited to change the configuration from \emph{'AXO'} to \emph{'BXO'}, which is also an allowed configuration. This peculiarity makes \emph{'AXO'} an "ambiguous" McIntosh configuration. An example of an unambiguous McIntosh configuration is \emph{'AX '} with a missing classification of the distribution of the AR in consideration. This was considered unambiguous because the distribution component is missing and only the \emph{'X'} class is allowed as the third component for this McIntosh configuration (as shown in Table \ref{table:2}), while the first two components followed an allowed configuration.
    
        After defining the different inconsistencies found in the records of McIntosh classifications, and following the sorting method based on the AR number, date, and time, an algorithm iterated through each entry with information about ARs for a given year. Because the datasets from each catalog were previously sorted according to the AR number, date, and time of observation, when the algorithm found an ambiguous entry, it inspected the most recent previous or next record of the same AR. If an acceptable McIntosh classification was found, referring to a neither ``ambiguous'' nor ``unambiguous'' McIntosh class, such a class replaced the McIntosh class in question, irrespective of which observatory the record was acquired. The unambiguous McIntosh configurations were found individually and corrected accordingly.
        
        In order to keep the data formatting consistent throughout the merged AR database, each AR record in the USAF catalogs had to be approximated to midnight, similar to the SWPC NOAA SRS records. Following the sorting method previously used for editing specific records of ARs, our algorithm iterated through each entry in the catalog for a given year and approximated its day and time to the next midnight based on four conditions: 
        \begin{enumerate}
            \item If the next entry has the same day and same AR number, store its index and continue to the next entry.
            \item If the next entry has the same day and a different AR number, approximate the current entry to midnight.
            \item If the next entry has the same AR as the next day, approximate the current entry to midnight.
            \item If the last entry is reached, approximate this entry to midnight.
        \end{enumerate}

        We approximate and keep only the last record of an AR during a day, independently from which observatory the record was noted. \emph{Condition 1} ensures that if the next AR record has the same date, meaning that such a record is not the last observation of the same AR during a given day, the corresponding index will be saved. Later, saved indices were dropped, keeping only the last and the approximated record of each respective AR during each day. When the first condition was not fulfilled, \emph{Condition 2} examined if the next AR record is from a different AR. If true, this meant that the current AR record was its last observation of the day, since AR records were sorted in ascending order of AR number, date, and time, and it was approximated to midnight. If previous conditions were not met, it meant that the next record had the same AR number with a different date. \emph{Condition 3} determined whether the next AR record corresponded to another day. If true, it meant that the current record was the last observation of the respective AR during the current day, and such an AR record was approximated to midnight. Because the last entry of the sorted catalogs for each corresponds to the last observation of any AR during that day, \emph{Condition 4} approximates such a record to midnight.

        Following the time approximation of the respective ARs, it was natural to also approximate their longitudinal location according to the Carrington rotation rate using the following formula:

        \begin{gather}
            \text{New Longitude} = \text{Current Longitude} + \left( \frac{\text{difference in hours until midnight}}{24\, {\rm hrs}} \right) *14.2^\circ
            \label{eqn:carringtonrot}
        \end{gather}
        
\begin{figure*}[!ht]
	\gridline{\fig{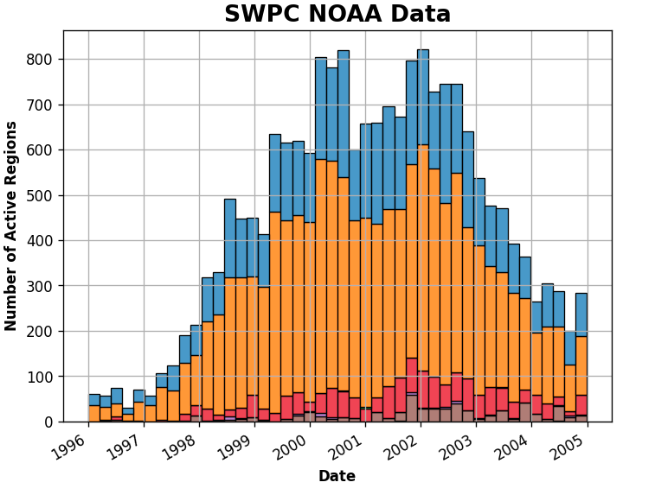}{ 0.39\textheight }{(a)}
        \fig{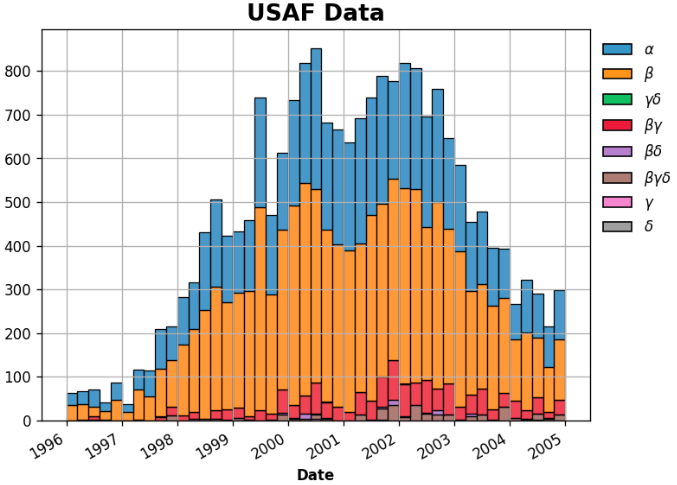}{0.40\textheight }{(b)}}
	\caption{Histograms used to perform the homogeneity test. The left panel (a) shows data originating from the Space Weather Prediction Center of the North Oceanic and Atmospheric Administration (SWPC NOAA). Right panel (b) shows data from the US Air Force (USAF). The legend shows the Hale classifications and the contribution of Active Regions (ARs) in a respective Hale class to the total number of ARs in each bin. The appearance of each class is consistent throughout the entire figure.}
	\label{fig:htest}
\end{figure*}

        By making the modifications described above, we were able to construct the continuous AR dataset by combining the data catalogs from the SWPC NOAA and USAF for December 1981 - February 2021. This dataset is presented in Figure \ref{fig:ARs}(a). We have also tested the homogeneity of the data provided by the two sources we utilize after applying the processing steps above to AFRL records. This was done through quantitative comparisons of the number of ARs and their respective features during several overlapping years, from January 1, 1997, to December 31, 2004, during the SC23 maximum. Figure \ref{fig:htest} shows the  histograms used for our homogeneity test. The histogram on the left was generated using data solely from the SWPC NOAA database, while the histogram on the right was produced using data from the USAF database. Although the total number of ARs in each dataset is relatively close, the height of several bins corresponding to a specific period of time seems to differ, meaning that the number of ARs recorded by the two solar centers for every single time is not exactly the same. In addition, the test showed that the portion of ARs with  \BGD \space (brown), \BG \space ARs (red), and \B \space (orange)  magnetic field types also slightly differ in these databases. Thus, we determined that the datasets obtained from the two different solar centers while generally consistent are not entirely homogeneous, and this has to be taken into account while utilizing the developed dataset for research purposes.

    \subsection{Linking Records of Solar Energetic Particle (SEPs) and Active Regions}\label{sec:SEPevent}

        The SWPC NOAA list of SEP events affecting the Earth provides records of the SEP events that, according to the observations of the Geostationary Operational Environmental Satellite (GOES), reached the threshold of 10 particle flux units (pfu) for $\ge$10\,MeV protons. The current records of SEP events in our possession span six decades, starting from April 1976 until September 2017. For the complete list of the SEP events please refer to the original source, \url{https://umbra.nascom.nasa.gov/SEP/}. The list contains information about the SEP event start and peak time, the peak flux of $\ge$10\,MeV protons, and the corresponding information about the preceding Coronal Mass Ejection (CME) and soft X-ray flare record. The majority of the records also contain information about the location of the host active regions on the solar surface and their NOAA number, allowing us to link this list with the homogeneous AR dataset constructed above. We also note here that, although the SEP records provide their start and peak date and time, we utilize the time of the preceding flare (more precisely, its peak in 1-8\,$\AA$ soft X-ray emission) as a reference time for merging the datasets, because the SEP arrival time may vary from minutes to several hours. Inconsistencies were also found in the SEP records and were updated to undertake this statistical analysis. The total number of SEP records in the analyzed list is currently 267. From that number, 36 records did not have the flare maximum date and time and were removed, leaving 231 SEPs.

\begin{figure*}[!ht]
	\gridline{\fig{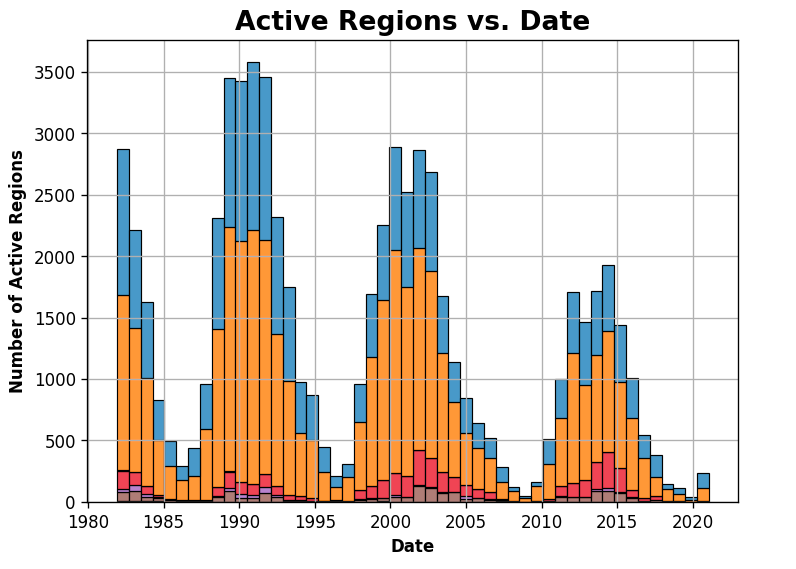}{ 0.46\textwidth }{(a)}
		\fig{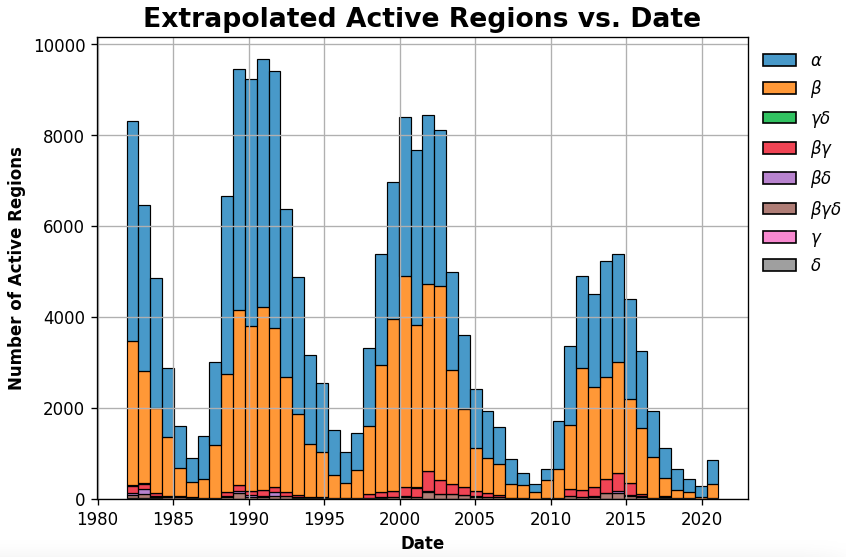}{0.50\textwidth}{(b)}
	}
	\caption{The continuous AR database was achieved by combining Active-Region (AR) catalogs from the Space Weather Prediction Center at the National Oceanic and Atmospheric Administration and US Air Force Space Weather Wing. The annual AR numbers are visualized through a stacked histogram in panel (a). Panel (b) shows a stacked histogram with the total number of extrapolated Active Regions vs. date. The legend shows the Hale classifications and the contribution of ARs in a respective Hale class to the total number of ARs in each bin. }
	\label{fig:ARs}	
\end{figure*}

        Not all SEPs originated from ARs observed on the visible solar disk, or within the $[-90^\circ, 90^\circ]$ longitude range. Figure \ref{fig:props}(a) shows that several SEP events were identified as originating from the ARs behind the western limb with a longitude $> 90^\circ$. The western hemisphere is more directly magnetically connected to the Earth \citep{Parker1958} and the statistical studies of SEP origins demonstrated the asymmetry towards the western limb \citep{Cliver2020}. Therefore, it is important to include information about the ARs located close to the western limb and to map some of the SEP events to these ARs. To accommodate for such SEPs, we decided to extrapolate the longitude of all ARs (and their corresponding dates) to cover the entire $360^\circ$ circumference of the Sun (i.e. until the ARs reach the Eastern limb again) by applying the Carrington rotation rate. In order to avoid any duplicates, the AR longitudes were extrapolated only after their last records near the west limb. The extrapolation was performed assuming the Carrington rotation rate following Eqn.~\ref{eqn:carringtonrot}. Figure \ref{fig:ARs}(b) illustrates the continuous AR database after performing the extrapolation. It can be interpreted as a stacked histogram with the total number of extrapolated Active Regions per year. The legend shows the Hale classifications and the contribution of Active Regions (ARs) of a respective Hale class to the total number of ARs in each bin, corresponding to a range of dates. It can be observed from each bin that the number of ARs increased approximately by a factor of three compared to the histogram of ARs in Figure \ref{fig:ARs}(a).

        After acquiring the extrapolated AR database and the SEP database, we merged the records into a single dataframe containing a one-to-one correspondence between ARs and SEPs generated by ARs. The merging process was performed by considering the corresponding AR numbers and the date of the consideration of an AR with the date of the peak flare time record from a SEP event (i.e., the class of the AR recorded in the preceding midnight with respect to the flare peak time was linked to the flare and, correspondingly, to the SEP records). From the 231 SEP events selected, we were able to match 181 SEPs to their AR sources in our extrapolated AR database.

\section{Results and Discussion} \label{sec:analysis}

    \subsection{Hale Classification}\label{sec:ResultsHale}

\begin{figure*}[!ht]
  \gridline{
  \fig{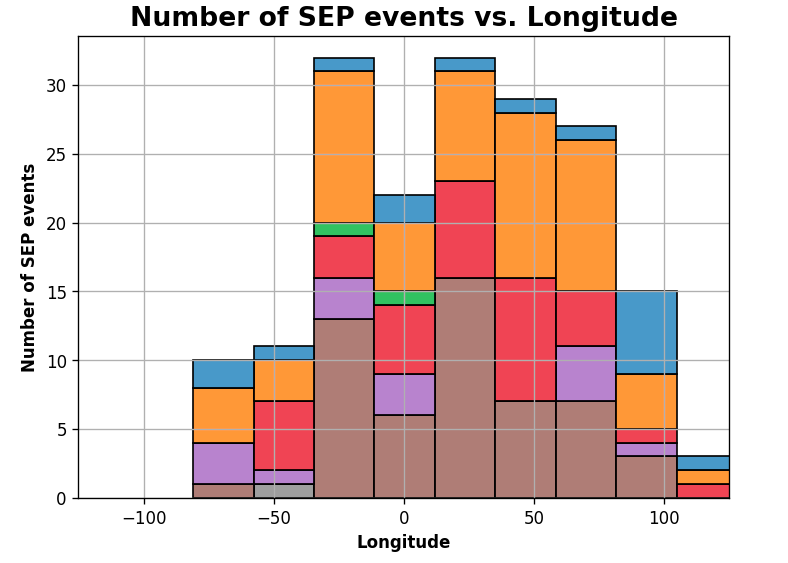}{ 0.50\textwidth }{(a)}
  \fig{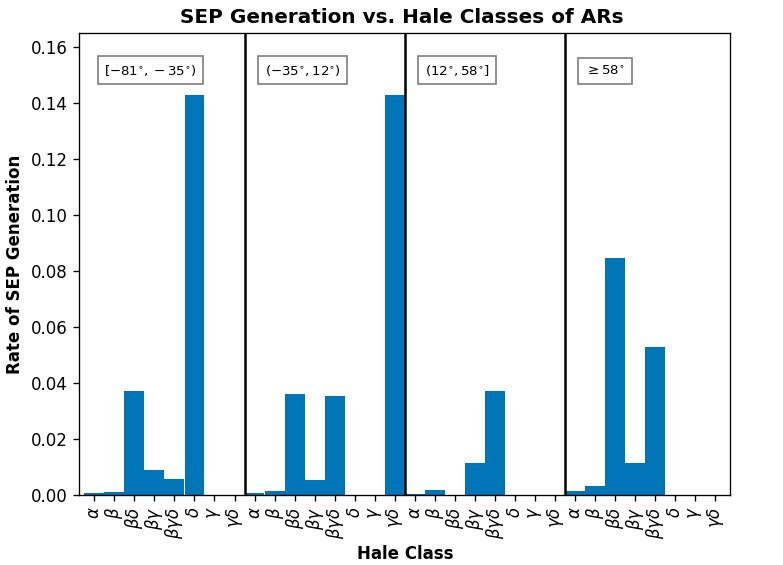}{ 0.49\textwidth }{(b)}}
  \gridline{
  \fig{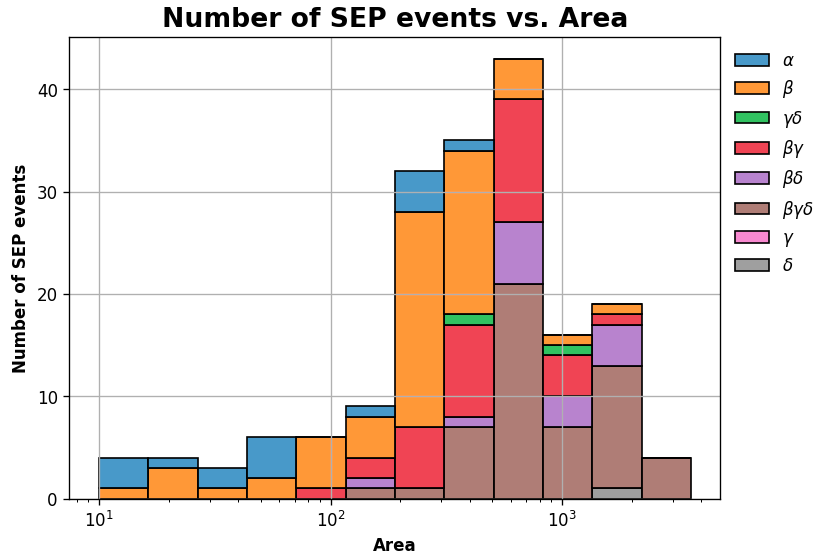}{ 0.40\textheight }{(c)}
  \fig{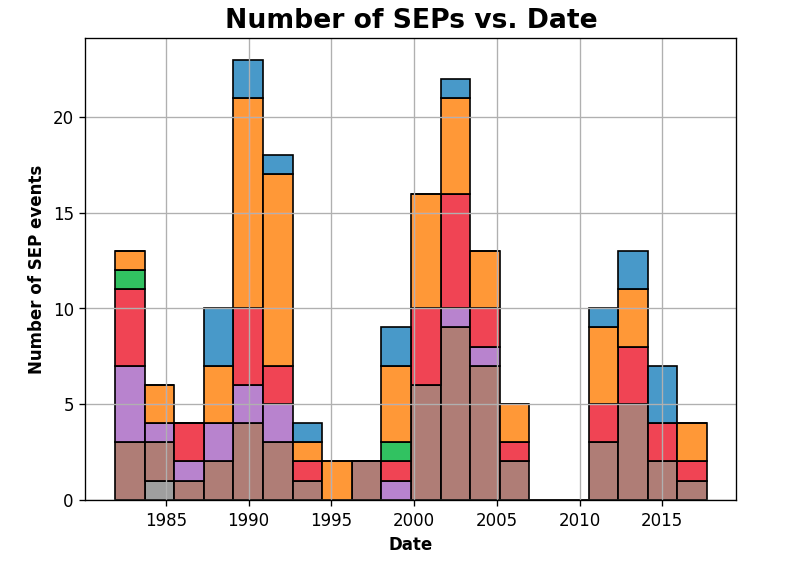}{ 0.39\textheight }{(d)}}
    \caption{Panel (a) shows the number of SEPs vs. the location of ARs along the surface of our Sun. According to this figure, the majority of SEPs were generated in the Western Hemisphere. The legend shows the Hale classifications and illustrates the contribution of ARs in a respective Hale class to the total number of SEPs in each bin. Panel (b) depicts the rate of SEPs produced by Active Regions (ARs) in certain longitude ranges classified by their magnetic field type. Panel (c) shows the number of SEPs vs. the area of their respective ARs. Panel (d) shows the number of SEPs within the specific date ranges in each bin. The legend shows the Hale classifications and illustrates the contribution of ARs in a respective Hale class to the total number of SEPs in each bin.}
    \label{fig:props}
\end{figure*}

        Figure~\ref{fig:props}(a) shows the total number of SEP events versus the longitude of the SEP-productive AR at midnight before initiating the SEP event. The legend shows the contribution of ARs with a respective Hale classification to the number of SEP events within each range of longitudes. Most SEPs are produced in the Western Hemisphere of the Sun, within the range $[0,90]$. This characteristic arises because the Earth is magnetically connected to the solar longitudes of $\approx{}75^\circ$, meaning that SEPs are more likely to reach Earth if it originates from an AR closer to that range of longitudes. Although this relationship was known and highlighted in the previous studies \citep[e.g.,][]{Cliver2020}, there are some additional interesting dependences related to the magnetic classes of SEP-productive ARs that we can mention.  Table~\ref{tab:tablehaleclass} presents the summary statistics of SEP-productive ARs depending on their Hale classes and summarizes the median locations in particular. One can notice from this table that the simpler configurations of Hale ARs (\A~and \B) have their median locations toward the western limb ($60^\circ$ and $30^\circ$, correspondingly) with respect to more complex \BG, \BD, and \BGD~regions ($29^\circ$, $-10^\circ$, and $24^\circ$). It can be also inferred from Figure~\ref{fig:props}(a) that a greater number of SEPs came from an \A \space (blue) AR when they were closer to the Earth and the Sun's magnetic connection, while the distribution of SEPs originating from \BGD \space (brown) ARs shows that a relatively large number of SEPs were generated far from $\approx{}75^\circ$. This can be related to the fact that more complex ARs generate stronger flares (in terms of their soft X-ray class) that statically result in faster and wider CMEs. The CME width was recently indicated to be an important parameter for SEP forecasting \citep{Torres2022}; on the other hand, the recent work by~\citet{Laitinen2023} indicated that SEPs can arrive at a wide range of longitudes, even without a wide particle source. We also notice that the AR records close to the western limb and behind it are extrapolated records and for them, the AR class is assumed to be unchanged from the last reliable observations. Correspondingly, we cannot exclude the possibility that SEP records from relatively simple \A~and \B~regions close to the western limb can correspond to more complex AR configurations evolved from these \A~and \B~regions. It also can be inferred that only six SEP events were found to be generated from behind the western limb.
        
        Figure~\ref{fig:props}(b) displays the rate of SEP event generation (i.e., a daily climatological probability of the SEP to be produced from the AR of a certain class) in four longitude bins: $(-81^\circ,-35^\circ)$, $(-35^\circ,12^\circ)$, $(12^\circ,-58^\circ)$, and $>{}58^\circ$. The daily climatological probability is calculated as the number of SEPs generated from ARs in a respective Hale class divided by the total number of ARs in our AR database with the same respective Hale class multiplied by 100\%, taking into account that each AR has only one record daily. One can notice that the probabilities increase towards the western longitudes for both the simpler configurations of ARs (like \B) and more complex \BGD, both categories of the regions become more SEP productive when approaching the magnetically-connected longitudes.
  
        Figure \ref{fig:props}(c) shows a histogram with the number of SEPs vs. the area of their respective ARs. It can be generally inferred that ARs with larger areas are more likely to produce SEP events. Table~\ref{tab:tablehaleclass} demonstrates that the median areas of SEP-productive ARs vary with their Hale class, and the same is evident in Figure \ref{fig:props}(c).
  
        Figure \ref{fig:props}(d) shows the number of SEPs events vs. their respective dates. Comparing with the histograms in Figure \ref{fig:ARs}, we can see that the SEP event productivity varies in phase with the 11-year solar activity cycle. It also demonstrates the known pattern of the solar cycle 24 being weaker than preceding cycles 23 and 22 both in terms of the sunspot numbers, numbers of ARs that appeared at the surface, and the number of generated SEPs.

\begin{figure*}[!ht]
\gridline{
   \fig{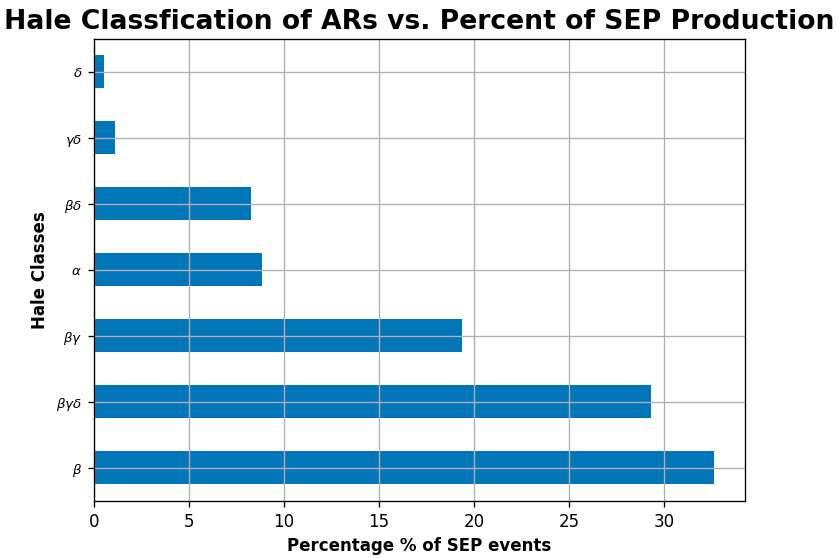}{ 0.49\textwidth }{(a)}
   \fig{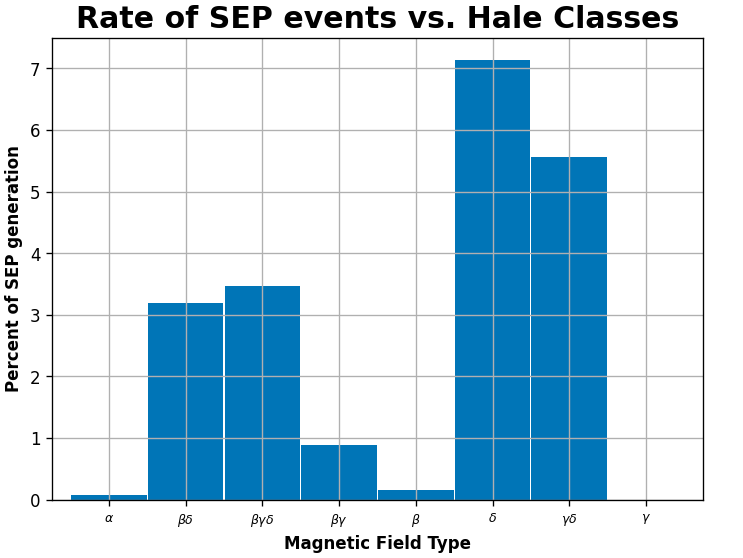}{ 0.43\textwidth }{(b)}
  }
  \caption{Panel (a) shows the fraction of the total number of SEP events (as a percentage) that were produced by ARs with a corresponding Hale class. Panel (b) shows the rate of SEPs produced by ARs with a corresponding Hale class. The rate is calculated by dividing the total number of SEPs produced by an AR from a respective Hale class and its corresponding total number of appearances.}
    \label{fig:Hale}
\end{figure*}

        Figure \ref{fig:Hale} shows the correlation between the Hale classification of ARs (described in Section \ref{sec:Hale}) and SEP productivity. Panel (a) shows the percentage of the total number of SEPs produced by ARs classified by a respective Hale class. The percentage was calculated by dividing the total number of SEP events found in a corresponding Hale class by the total number of SEPs in our database, which is 181 SEP events in total. Approximately 33\% of the total number of SEPs were produced by \B \space ARs, followed by the \BGD \space class with approximately 29\%. The \D~ARs appear to have produced only one SEP event, and \GD~ have produced only two of them (see Table~\ref{tab:tablehaleclass}). Interestingly, even the ARs that are typically assumed to be not active with respect to solar transient events (such as \A~or~\B) generated more than 40\% of the SEPs considered in this work if combined. This confirms the importance of the characterization of ARs with other parameters in addition to the Hale class, such as McIntosh classes or quantitative magnetic field properties~\citep{Kasapis2022,Sadykov2021}.

        The ARs exhibiting the \D \space Hale class in their magnetic field configuration (which are \D , \BD , \GD , and \BGD ) are historically regarded as highly SEP productive because of their correlation with solar flares. Figure \ref{fig:Hale}(b) shows the rate of SEP productions (i.e., the daily climatological probability of SEP events in these regions, shown as a percentage on the y-axis) of ARs in a respective Hale class. It can be inferred that the \BGD \space configurations have the high chance of the SEP event generation (approximately 3.5\%) given the condition that the \BGD \space region is observed, confirming what has been historically concluded. The \BD \space regions show a rate of approximately 3.2\%. Although the rates of the \GD \space ARs (5.6\%) and \D \space ARs (7.1\%) are higher with respect to \BGD \space regions, we note that only 2 and 1 SEP event records, correspondingly, exist in our merged data set, and these rates should be interpreted with caution.

    \subsection{McIntosh Classification}\label{sec:ResultsMcIntosh}
    
\begin{figure*}[!ht]
\gridline{\fig{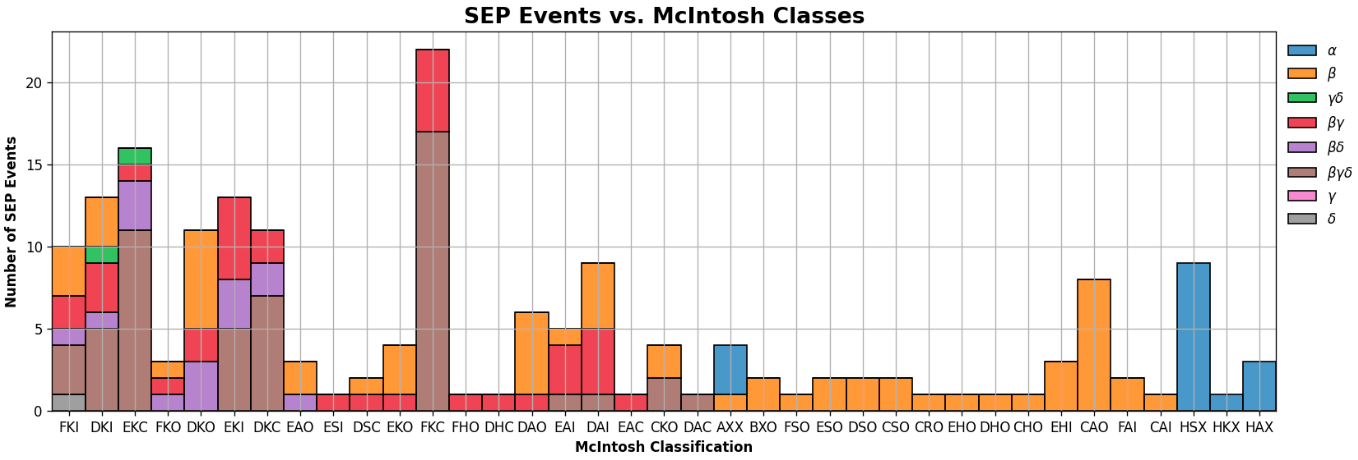}{0.99\textwidth}{(a)}
          }
\gridline{\fig{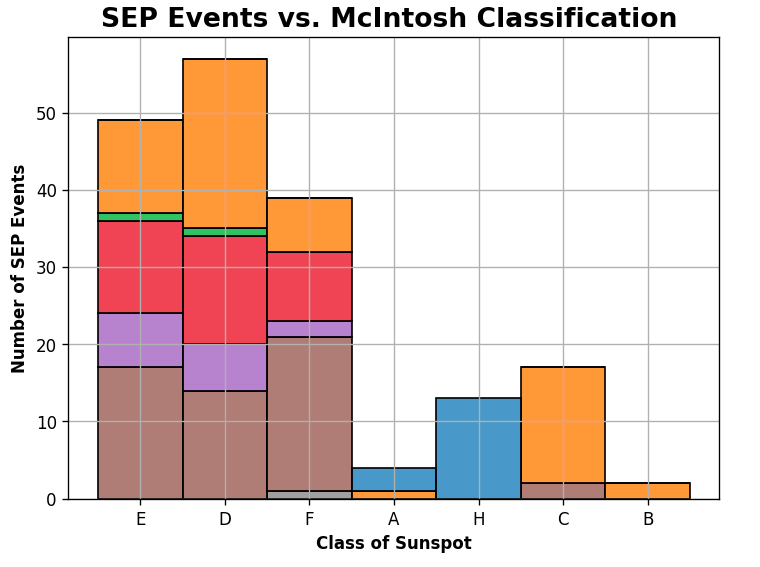}{0.45\textwidth}{(b)}
          \fig{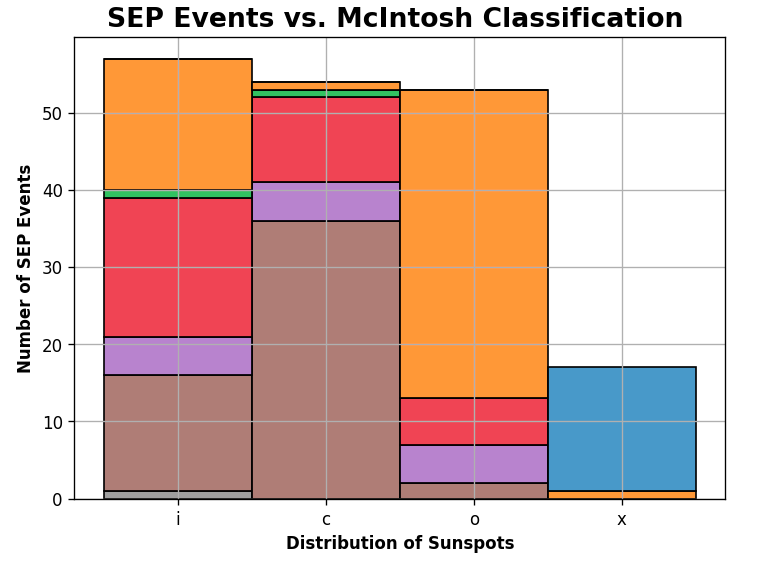}{0.46\textwidth}{(c)}}
    \caption{Panel (a) shows the distribution of the SEP events over the entire McIntosh Classification scheme. The legend shows the Hale classifications and the contribution of Active Regions (ARs) in a respective Hale class to the total number of SEPs in each bin. The appearance of Hale's class is consistent throughout the entire figure. After separating each of the McIntosh components, panel (b) shows the number of SEPs produced by ARs classified by the first component of the McIntosh scheme. Panel (c) shows the number of SEPs produced by ARs classified by the third component of the McIntosh component.}
    \label{fig:McIntosh13}
\end{figure*}

        Figure \ref{fig:McIntosh13} shows the results of a statistical study referencing the McIntosh classification of ARs, described in Section \ref{sec:McI}, and SEP events. Panel (a) shows the distribution of the total number of SEPs that were produced by ARs with their respective three-component McIntosh classification. Through the legend, we can distinguish the proportion of ARs from a corresponding Hale class that produced SEPs. Thus, it can be inferred that the largest quantity of SEP events originated from ARs having 'k' as the second McIntosh component. Additionally, the majority of these ARs were classified as \BGD, which allows us to close in on a connection between the Hale class (\BGD) and the McIntosh subclass (k). Figure \ref{fig:McIntosh13}(b) shows the distribution of SEP events generated from ARs with a respective AR class given by the first component of the McIntosh classification scheme, while Figure \ref{fig:McIntosh13}(c) indicates the total number of SEPs which originated from ARs with a respective distribution of Sunspots provided by the first McIntosh component. Because the legend is kept consistent throughout the entire figure, it can be inferred from Figure \ref{fig:McIntosh13}(b) that most SEPs were generated from ARs exhibiting classes E, D, F and corresponding Hale classes \BGD, \BD, \BG, and $\beta$. On the other hand, Figure \ref{fig:McIntosh13}(c) shows an almost even total number of SEPs originated from ARs with the distribution of sunspots $i$, $c$, and $o$ with a less even proportion of such ARs in a respective Hale class. The two most obvious correlations are the \BGD \space regions, which appear to favor the 'c' distribution of sunspots, and the $\beta$ regions favoring the 'o' counterpart.

\begin{figure*}[!ht]
\gridline{
   \fig{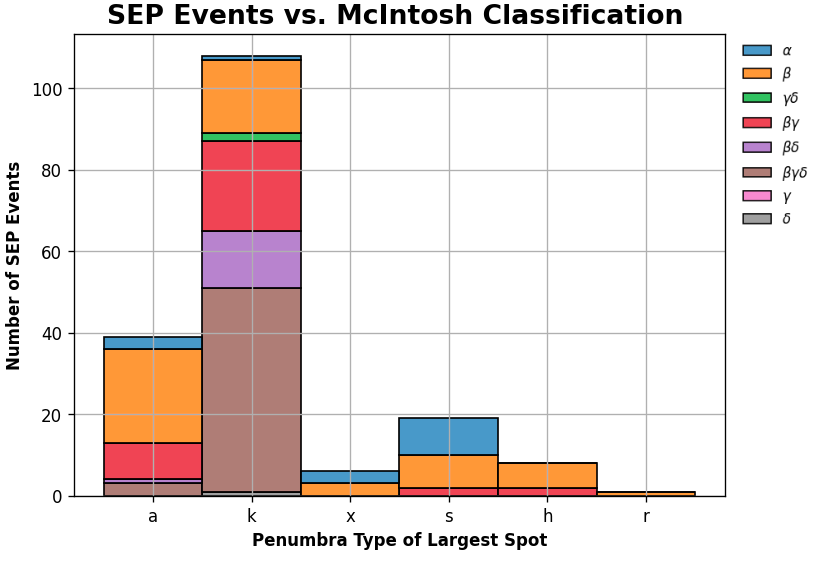}{0.50\textwidth}{(a)}    
    \fig{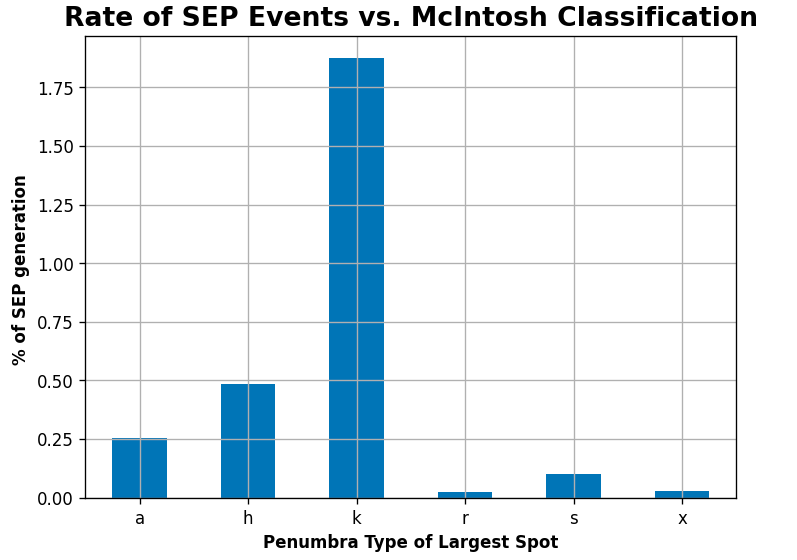}{0.48\textwidth}{(b)} 
        }
    \caption{Panel (a) shows the number of SEP events produced by ARs classified by the second component of the McIntosh classification scheme. The legend shows the Hale classifications and the contribution of Active Regions (ARs) in a respective Hale class to the total number of SEPs in each bin. Panel (b) shows the rate of SEP production of ARs classified by considering the McIntosh component used in panel (a). To find calculate the rate, the total number of SEPs generated from ARs with a respective second McIntosh component is divided by the total number of appearances of ARs with the same corresponding classification.  }
    \label{fig:McIntosh2}
\end{figure*}

        A more defined association of SEP events and McIntosh classes is shown in Figure \ref{fig:McIntosh2}(a). It shows the total number of SEP events generated by ARs with a given penumbra type of the largest sunspot in the AR, defined as the second component of a McIntosh class. Based on this figure, we can conclude that ARs with a 'k' subclass as the second component of a McIntosh classification is the most SEP productive. In addition, it can be inferred that ARs with a \BG, \BD, \GD, and \BGD \space magnetic-field configurations favor a 'k' subclass as the second component of a McIntosh class in terms of SEP production. Figure \ref{fig:McIntosh2}(b) shows the rate of production of SEPs from ARs with a respective penumbra type of their corresponding largest sunspot. The rate was calculated as calculated in Section \ref{sec:ResultsHale} for Hale classes. The number of SEPs generated from ARs in a respective class is divided by the total number of ARs in our AR database with the same respective class multiplied by 100\%. Following the results, it can be concluded that ARs with the 'k' subclass as the second McIntosh component has a higher rate of SEP production, and consequently, they are more probable to produce SEPs than ARs with any other class in this component.

    \subsection{Combined McIntosh and Hale Classifications}

\begin{figure*}[!ht]
    \gridline{
   \fig{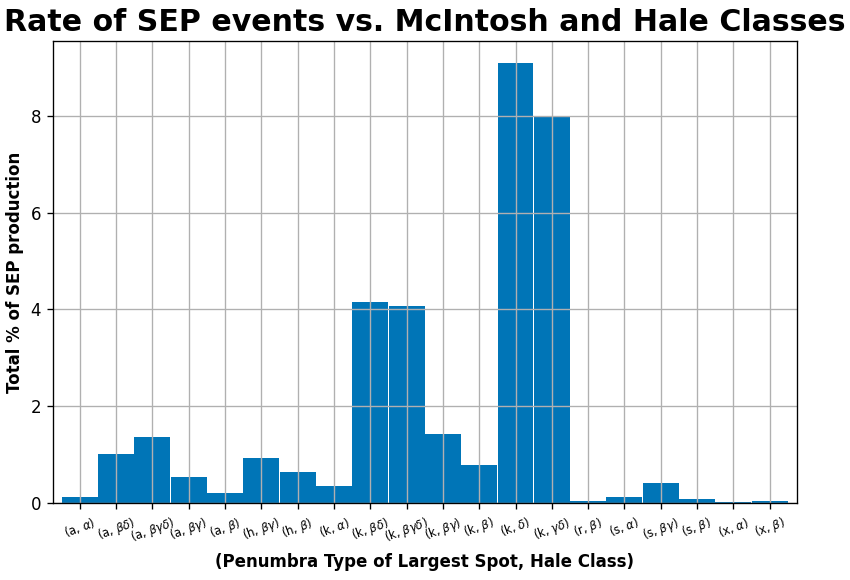}{ 0.70\textwidth }{}
    }
    \caption{Rate of SEP events (in percentage) vs. a combination of the second component of the McIntosh classification scheme and the Hale classification scheme. When combined, the rates of SEP production are found to increase for either or both classes in consideration when considered separately. Two major groups of ARs are found to be the most SEP productive: (k, \D) and (k, \GD) ARs.  }
    \label{fig:ResultsHaleMcIntosh}
\end{figure*}

        Figure \ref{fig:McIntosh2} shows that some ARs are more likely to produce SEPs when their second McIntosh component is determined as 'k' and paired with certain Hale class configurations. To further study this relationship, we combined the two classification schemes and ARs were grouped based on two features: the Hale class and the second component of the McIntosh class. The results are shown in Figure \ref{fig:ResultsHaleMcIntosh}. We can conclude that the rate of SEP production is the highest for ARs with any given Hale class combined with the 'k' subclass as the second component of a McIntosh classification. In addition, it is important to note that these rates are higher than the rates found by separately considering Hale classes and the considered subclass of the McIntosh classification. For example, the rates of SEP events produced by \BGD \space ARs and ARs with 'k' second McIntosh components were $\sim 3.5\%$ and $\sim 1.8\%$ in that order and as shown in Figures \ref{fig:Hale}(b) and \ref{fig:McIntosh2}(b). Because (k, \BGD) ARs have an $\sim 4.1\%$ rate of SEP production, they are more likely to produce SEPs than the ARs that fall only into one of the 'k' or \BGD~class categories. The same can be concluded for \B, \BD, \BG, \GD, and \D \space ARs if they are also determined to have the 'k' McIntosh subclass. These results indicate the importance of including both Hale and McIntosh classes for advancing the forecasting of SEPs.

\section{Summary} \label{sec:summary}
  
    We presented a statistical study of the correlation between SEP events and properties, Hale, and McIntosh classes of ARs. Properties of ARs included longitude and area of ARs, while classes came from the Hale and McIntosh classification schemes. We concluded that the longitudes closer to the magnetic connection between the Sun and the Earth are more favorable for less complex ARs, such as \A \space and \B \space ARs, in terms of SEP production. ARs of larger areas, irrelevantly to their Hale class, are also more likely to produce SEP events. The total number of SEPs produced by ARs of a certain Hale or McIntosh class is not a sufficient characteristic to determine the hazard a given AR may represent. For instance, although \BD \space ARs generated only 15 SEP events throughout the entire time frame in consideration, they have a higher rate of producing SEP events (about 3.2\%) with respect to \B~ regions that produced the most number of SEPs (59). It was found that most SEPs were generated from an AR classified with the 'k' McIntosh subclass as the second component ($\approx$60\%, or 108 out of 181 considered events), and some of these ARs are more likely to produce SEPs if they fall in a certain Hale class. For example, \BGD \space ARs have a $\sim 3.5\%$ chance to produce SEPs, while ARs with a 'k' as the second McIntosh component have a $\sim 1.8\%$ rate of SEP production. An AR with a combination of these two classes has a probability of $\sim 4.1\%$ to produce a SEP event. The developed homogeneous dataset of AR classes spanning more than three solar cycles (1981-2021) can be utilized for the development of robust SEP events and other transient solar activity forecasting approaches validated over an extensive time period and varying solar activity.

\section*{\textsc{Acknowledgments}} \noindent \centering \normalsize This research was supported by NASA Early Stage Innovation program grant 80NSSC20K0302, NASA LWS grant 80NSSC19K0068, NSF EarthCube grant 1639683, and NSF grant 1835958. VMS acknowledges the NSF FDSS grant 1936361 and NSF grant 1835958. EI acknowledges the RSF grant 20-72-00106.

\noindent
\begin{table}[h]
    \begin{tabular}{| c | c | c | c | c |}
        \hline
        Hale  &   \# of SEP  &   Median Longitude    &   Median AR area, &   Median peak flux    \\
        Class  &  events   &    of SEP-active ARs\footnote{Determiner as the median longitude of the SEP-active ARs at midnight before initiating the SEP event.}, deg   &   $10^{-6}$th hemispheres &   of SEPs, pfu    \\
        \hline
        \A  &   16    &   60.0+/-37.5    &    60.0+/-50.0   &   91.5+/-75.5   \\
        \hline
        \B  &   59    &   30.0+/-42.0    &   240.0+/-130.0  &   55.0+/-41.0    \\
        \hline
        \BG  &   35    &   29.0+/-27.0   &    450.0+/-210.0    &   66.0+/-51.0   \\
        \hline
        \G  &   0    &   -   & -   &   -    \\
        \hline
        \BD  &   15    &   -10.0+/-49.0    &    820.0+/-300.0   &    180.0+/-162.0   \\
        \hline
        \BGD  &   53    &   24.0+/-32.0   &    780.0+/-310.0   &    134.0+/-120.0   \\
        \hline
        \GD  &   2    &   -12.0+/-22.0    & 745.0+/-405.0   &   90.0+/-60.0    \\
        \hline
        \D  &   1    &   -43.0+/-0.0   &   1870.0+/-0.0    &   2500.0+/-0.0    \\
        \hline
    \end{tabular}
    \caption{Properties of the SEP-productive active regions (ARs) and the corresponding SEP events depending on their Hale class. Notice that \GD~and \D~have poor statistics of SEP events produced from those regions (2 and 1 SEP event, correspondingly) and there are no SEP events associated with \G~regions in our database.}
    \label{tab:tablehaleclass}
\end{table}

\clearpage

\bibliography{HaleAndMcIntoshAnalysis.bib}
\bibliographystyle{aasjournal}



\end{document}